\documentclass[onecolumn]{revtex4}
\usepackage{amsmath,amssymb,graphics,epsfig,subfigure}
\usepackage{color}
\usepackage[colorlinks,linkcolor=red,anchorcolor=red,citecolor=green]{hyperref}
\usepackage{setspace}
\usepackage{booktabs}
\usepackage{amsfonts}
\usepackage{bbm}
\usepackage{graphicx}
\usepackage{pgfplots}
\usepackage{mathrsfs}
\usepackage{color}
\usepackage{verbatim}
\usepackage{cancel}
\usepackage{float}
\begin{document}
	\thispagestyle{empty}
\begin{center}
	\title{Complex Plane Phase Diagram and Widom Line for the Born-Infeld Black Holes with Reentrant Phase Transition}
	
	\author{Fei Guo$^{1}$ and Zhen-Ming Xu$^{1,2,3,4}$
	\vspace{6pt}\\}

\affiliation{$^{1}$School of Physics, Northwest University, Xi'an 710127, China\\
	$^{2}$Peng Huanwu Center for Fundamental Theory, Xi'an 710127, China\\
	$^{3}$Shaanxi Key Laboratory for Theoretical Physics Frontiers, Xi'an 710127, China\\
	$^{4}$Fundamental Discipline Research Center for Quantum Science and technology of Shaanxi Province, China}
	
	\begin{abstract}
The Lee-Yang phase transition theory applied in the anti-de Sitter (AdS) black hole has inspired the exploration of complex phase diagram and supercritical phenomena in black hole thermodynamics. In this study, we extend the approach to the four dimensional Born-Infeld AdS black hole. This system exhibits a rich phase structure, including reentrant phase transitions, due to the modulation of the Born-Infeld nonlinear parameter. Through the Lee-Yang zeros, we obtained the complex phase diagram of the Born-Infeld AdS black hole and derived the supercritical crossover line---Widom line, which strictly originates from the first-order stable critical point. The results indicate that Born-Infeld nonlinear effects significantly alter the types and characteristics of phase transition in critical region, while do not disrupt the uniqueness of the Widom line in supercritical region. Our study uncovers a universal simplified feature of the thermodynamic behavior of nonlinear gravitational systems in supercritical region. It also deepens our understanding of the fundamental connection between critical phenomena and continuous phase transitions in the extended phase space of black holes.
	\end{abstract}

\maketitle
\end{center}

\section{Introduction}

Reentrant phase transition is an interesting form of non-monotonic thermodynamic behavior. In standard thermodynamic processes, a system typically evolves from an ordered phase to a disordered phase as the temperature increases (e.g., from liquid state to gas state). However, in reentrant phase transition systems, monotonic tuning of thermodynamic parameters (temperature or pressure) induces multiple phase changes, eventually reverting the system to its initial macroscopic phase. This concept originally arose from studies of phase equilibrium in multicomponent fluid mixtures. The classic “water-nicotine” system demonstrates a non-monotonic evolutionary sequence of single-phase mixing $\rightarrow$ two-phase separation $\rightarrow$ single-phase mixing~\cite{NARAYANAN1994135,SINGH2000107}. This phenomenon indicates that, in certain classical thermodynamic systems and complex fluids, the competition between entropy and intermolecular interactions can lead to phase diagrams with highly rich and multi-structured characteristics, rather than simple monotonic evolution.
			
With the deepening of research into black hole thermodynamics, people have discovered a fascinating duality between gravitational systems and classical fluids \cite{Maldacena:1997re,Chamblin1999,Kastor2009,Dolan2011,Kubiznak:2012wp,Kubiznak:2016qmn,Mann2025}. In the extended phase space, a negative cosmological constant is assigned a physical meaning as the thermodynamic pressure, with its conjugate quantity corresponding to the thermodynamic volume. Based on this framework, the charged AdS black hole has been shown to exhibit phase transition behavior highly similar to that of van der Waals fluids \cite{Kubiznak:2012wp,Kubiznak:2016qmn,Mann2025,Wei2022a,Wei2023,Karch2015,Mancilla2025,Zeng:2015wtt,Zeng:2015tfj}. Even more excitingly, black hole systems also display reentrant phase transitions: as the temperature decreases under constant pressure, the black hole system can undergo a sequential phase transition from “large black hole (LBH) $\to$ small black hole (SBH) $\to$ large black hole (LBH)”. This characteristic behaviour was first observed precisely in the Born-Infeld (BI) settings in~\cite{Gunasekaran:2012dq}. Since then, similar phenomena have been observed in higher-dimensional ($d\ge 6$) rotating AdS black holes~\cite{Altamirano:2013ane} and various Lovelock gravity models~\cite{Frassino:2014pha}.  The BI nonlinear term breaks the linear constraints of Maxwellian electrodynamics~\cite{Altamirano:2014tva,Breton:2004qa} and this nonlinear effect plays a thermodynamic role similar to high-dimensional gravitational corrections, enabling four-dimensional BI-AdS black holes to exhibit phase structures highly sensitive to the parameter $b$ ---including no phase transition, reentrant phase transitions, and standard first-order phase transitions~\cite{Chemissany:2008fy,Banerjee:2011cz,Banerjee:2012zm,Miskovic:2008ck,Fernando:2006gh,HENDI2015157,Hendi2016a,Hendi2016b,Aghbolagh2025,Dey:2004yt,Zou:2013owa}. This makes BI-AdS black hole an ideal theoretical model for exploring the impact of nonlinear effects on the thermodynamic structure of spacetime.
			
Current research on black hole systems exhibiting rich phase transitions primarily focuses on the region below the critical point, while the supercritical regime has received less attention. Taking supercritical carbon dioxide as an example, although no distinct gas-liquid coexistence exists in the supercritical regime, both experiments and simulations reveal significant fluctuations in kinetic and thermodynamic properties. The Widom line divides the supercritical region into ``liquidlike'' and ``gaslike'' zones, corresponding to the extrema of thermodynamic response functions such as the isobaric heat capacity~\cite{Simeoni2010}. Based on the duality between black hole thermodynamics and conventional fluid systems and due to the unique behavior of black holes, different schemes for defining the Widom line have been introduced in the thermodynamic system of black holes~\cite{Xu:2025jrk,Zhao2025,Li2025,Wang2025,Anand2025,Li2026}. However, considering that the relevant thermodynamic response function in the black hole thermodynamic system only exhibits local extremum characteristics in the supercritical region, the definition of the conventional Widom line becomes ambiguous. Recent work has introduced the Lee-Yang phase transition theory into black hole thermodynamics to provide a novel analytical framework for the problem~\cite{Xu:2025jrk}. The theory has been successfully applied to four-dimensional charged AdS black holes, yielding significant results.
			
Inspired by the study \cite{Xu:2025jrk}, we apply the Lee-Yang phase transition theory to four-dimensional BI-AdS black holes and investigate whether a well-defined Widom line persists under rich phase behaviors, and how it relates to the critical points. Specifically, by adjusting the nonlinear parameter~$b$, we consider three representative scenarios, including standard first-order and reentrant phase transitions. For each case, we construct the corresponding singularity maps in the complex~$r_h$ plane to analyze the supercritical thermodynamic behavior. This study not only confirms the applicability of the Lee--Yang phase transition theory in nonlinear gravitational systems, but also shows that a single Widom line structure emerges in the supercritical regime even for black hole models with rich phase transitions. The results provide a new perspective for understanding the profound connection between phase transition and critical phenomena in the extended phase space.

\section{Thermodynamics and Phase Structure}\label{II}
This section presents the thermodynamic foundation and phase structure of four-dimensional BI-AdS black holes. In Sec.~\ref{2.1}, we begin with the action of Einstein-Born-Infeld theory and derive the spherically symmetric solution in Schwarzschild coordinates. We then list some thermodynamic quantities, including the mass, temperature, entropy, electric potential, Gibbs free energy, and heat capacity. In Sec.~\ref{2.2}, the nonlinear parameter $b$ is divided into three intervals based on an analysis of the equation of state, each corresponding to a distinct type of phase transition. Phase diagrams are plotted in the $P-T$ and $G-T$ planes to illustrate and compare the characteristic features of these different phase transitions.
		
\subsection{Thermodynamics}\label{2.1}	
In the AdS spacetime, the action of the Einstein-Born-Infeld theory can be written as~\cite{Miskovic:2008ck,Fernando:2006gh,HENDI2015157,Hendi2016a,Hendi2016b,Aghbolagh2025}
\begin{equation}
I_{\text{EM}} = -\frac{1}{16\pi} \int_{M} \sqrt{-g} \left( R + \mathcal{L}_{\text{BI}} - 2\Lambda \right),
\end{equation}
where $\Lambda$ denotes the cosmological constant and the expression for the BI term is given by
\begin{equation}
\mathcal{L}_{\text{BI}} = 4b^2 \left( 1 - \sqrt{1 + \frac{1}{2b^2} F^{\mu\nu} F_{\mu\nu}} \right),
\end{equation}
the parameter $b$ represents the maximum electromagnetic field strength, which determines the nonlinear dynamical behavior of the electromagnetic field.
		
The spherically symmetric black hole solution and electromagnetic field are
\begin{equation}
ds^2 = -f(r)dt^2 + \frac{dr^2}{f(r)} + r^2(d\theta^2 + \sin^2\theta d\phi^2),
\end{equation}
and
\begin{equation}
F = \frac{Q}{\sqrt{r^4 + Q^2/b^2}} dt \wedge dr,
\end{equation}
where the metric function $f(r)$ is given by~\cite{Fernando:2003tz}
\begin{align}
f(r)= 1 - \frac{2M}{r} - \frac{\Lambda r^2}{3} + \frac{2b^2 r^2}{3} \left( 1 - \sqrt{1 + \frac{Q^2}{b^2 r^4}} \right) + \frac{4Q^2}{3r^2} \, _2F_1\left( \frac{1}{4}, \frac{1}{2}; \frac{5}{4}; -\frac{Q^2}{b^2 r^4} \right),
\end{align}
here $M$ and $Q$ correspond to the mass and charge of the black hole, respectively, and ${}_2F_1$ is the hypergeometric function. In the extended phase space, the cosmological constant is regarded as the thermodynamic pressure and treated as an independent thermodynamic variable, with the relationship given by
\begin{equation}
P = -\frac{\Lambda}{8\pi}.
\end{equation}
Hence the first law of thermodynamics is in terms of the event horizon radius $r_h$ which is the largest root of the metric function $f(r)=0$~\cite{Gunasekaran:2012dq},
\begin{equation}
dM = TdS + VdP + \Phi dQ + Bdb,
\end{equation}
where the thermodynamic volume $V=4\pi r_h^3/3$ and other related thermodynamic quantities are (to simplify the expression, we replace $\sqrt{1 + \frac{Q^2}{b^2 r_h^4}}$ with the letter $A$)
\begin{align}
M &= \frac{r_h}{2} + \frac{4}{3}\pi P r_h^3 + \frac{b^2 r_h^3}{3} \left( 1 - A \right) + \frac{2Q^2}{3r_h^2} \,_2F_1\left( \frac{1}{4}, \frac{1}{2}; \frac{5}{4}; -\frac{Q^2}{b^2 r_h^4} \right), \label{eq:M} \\
T &= \frac{1}{4\pi r_h} \left( 1 + 8\pi P r_h^2 + 2b^2 r_h^2 \left( 1 - A \right) \right), \label{teq} \\
S &= \pi r_h^2, \label{eq:S} \\
\Phi &= \frac{Q}{r_h} \,_2F_1\left( \frac{1}{4}, \frac{1}{2}; \frac{5}{4}; -\frac{Q^2}{b^2 r_h^4} \right), \label{eq:Phi} \\
B &= \frac{2}{3}b r_h^3 \left( 1 - A \right) + \frac{Q^2 }{3 b r_h}\,_2F_1\left( \frac{1}{4}, \frac{1}{2}; \frac{5}{4}; -\frac{Q^2}{b^2 r_h^4} \right). \label{eq:B}
\end{align}
According to the standard definition of thermodynamics, the Gibbs free energy and heat capacity at constant pressure can be obtained
\begin{align}
G&= \frac{1}{4}r_h - \frac{2\pi}{3} P r_h^3 - \frac{b^2 r_h^3}{6} \left( 1 - A \right) + \frac{2Q^2}{3r_h} \,_2F_1\left( \frac{1}{4}, \frac{1}{2}; \frac{5}{4}; -\frac{Q^2}{b^2 r_h^4}\right),\label{eq:fe} \\
C_P &= \frac{2\pi A r_h^4 -4\pi Q^2 r_h^2 + 16\pi^2 P A r_h^6 + 4\pi b^2 A r_h^6 - 4\pi b^2 r_h^6 }{2Q^2 - 2b^2 r_h^4 - A r_h^2 + 2b^2 A r_h^4 + 8P\pi A r_h^4}.\label{eq:Cp}
\end{align}

By rescaling the thermodynamics via the black hole charge, we have
\begin{equation}
r_{\mathit{h}} \propto |Q|,\quad S \propto |Q|^2,\quad T \propto |Q|^{-1},\quad P \propto |Q|^0,\quad G \propto |Q|,\quad C_P \propto |Q|^2.
\end{equation}
The initial thermodynamic functions can be expressed as forms that depend only on the dimensionless parameter $b$. According to the scale invariance of the thermodynamic system, the physical behavior remains unchanged under this transformation. Therefore, in the subsequent analysis, we treat the charge as a fixed parameter (setting $Q=1$) and focus on the influence of the nonlinear parameter $b$ on the Widom line in the supercritical region and the smooth crossover of thermodynamic response functions. This treatment does not compromise any physical rationality.

\subsection{Phase Structure}	\label{2.2}	
According to Eq.~(\ref{teq}), we derive the equation of state of the black hole \cite{Gunasekaran:2012dq}
\begin{equation}
P = \frac{T}{v} - \frac{1}{2\pi v^2} - \frac{b^2}{4\pi} \left( 1 - \sqrt{1 + \frac{16}{b^2 v^4}} \right),
\end{equation}
where the specific volume $v=2r_h$, and the critical point satisfies the following equations
\begin{equation}
\left. \frac{\partial P}{\partial v} \right|_T = 0,\quad \left. \frac{\partial^2 P}{\partial v^2} \right|_T = 0.\label{eq:critical value}
\end{equation}
According to reference \cite{Altamirano:2014tva,Xu:2019yub}, the parameter $b$ can be divided into three cases, corresponding to the no-phase-transition region, the reentrant phase-transition region, and the standard first-order phase-transition region.

{\bf Non-phase transition regime ($b < 0.3536$)} \quad In this parameter range,the system exhibits the simplest thermodynamic behavior. It is analogous to that of the Schwarzschild-AdS black hole, with no critical points(by solving the Eq.~(\ref{eq:critical value})) or phase transition phenomena present.
		
\begin{figure}[htbp]
	\centering
		\includegraphics[width=70 mm]{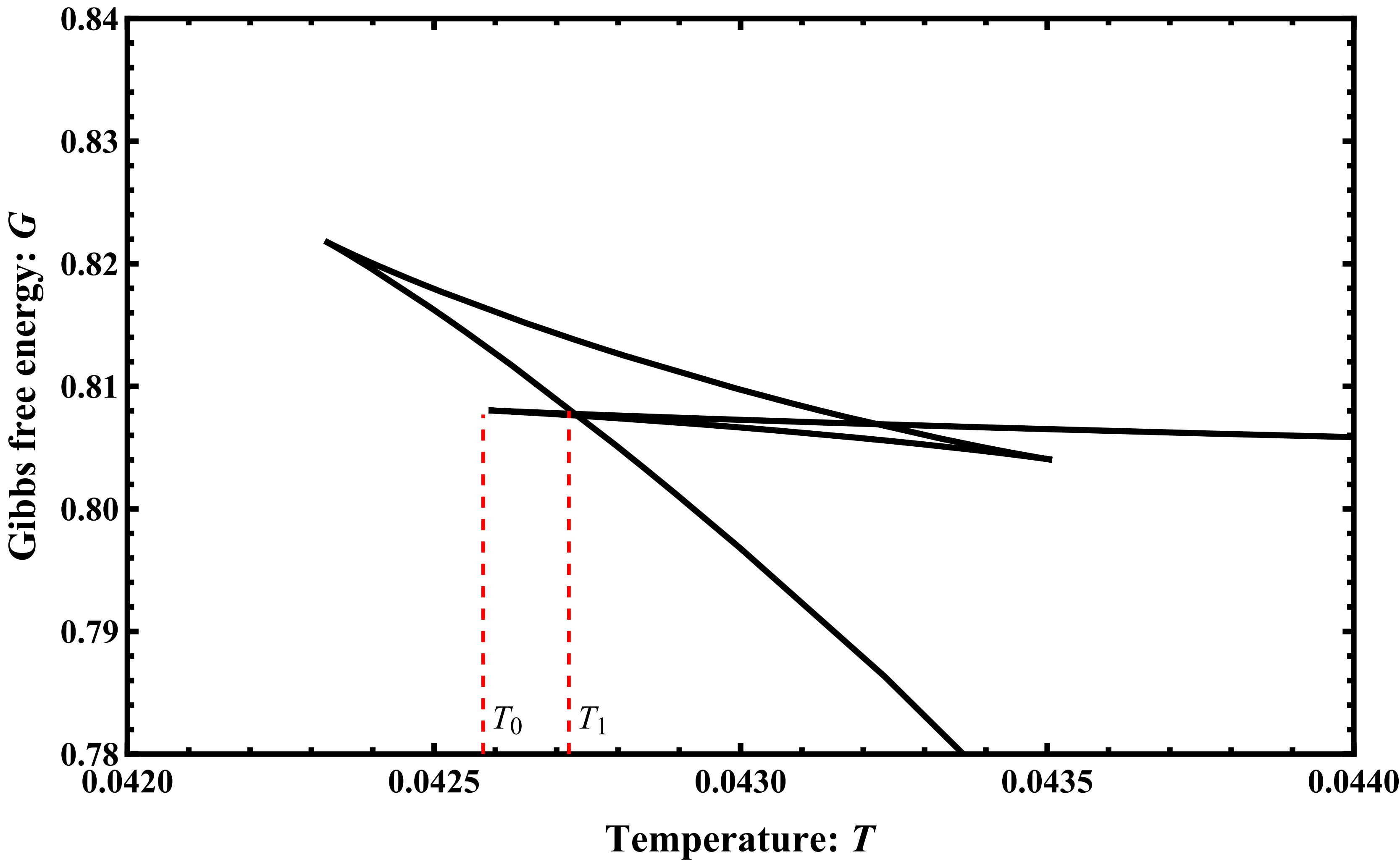}
        \includegraphics[width=70 mm]{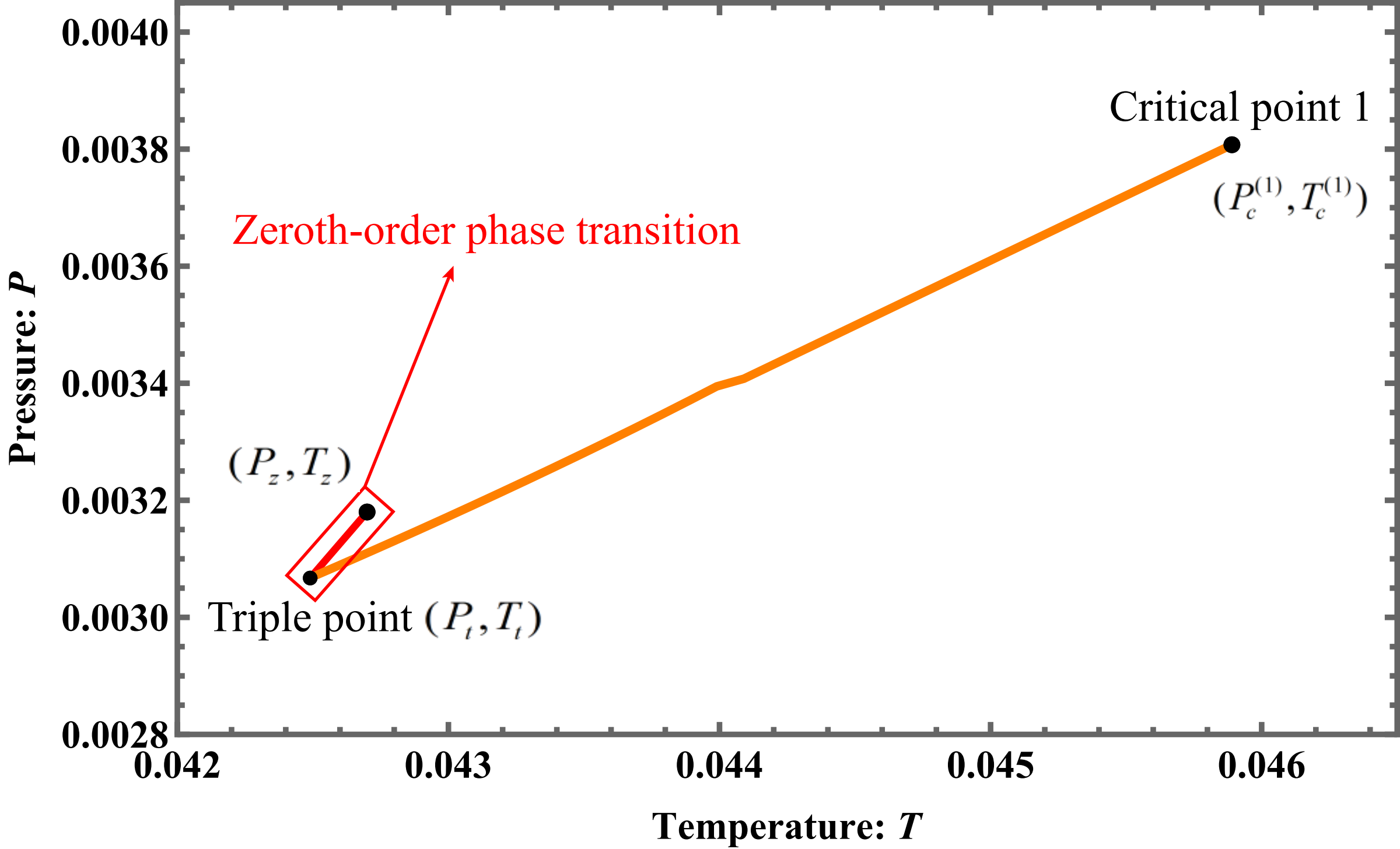}
	\caption{Phase diagrams of BI-AdS black hole with $b=0.4$ and in the left diagram we set $P=0.98P_z$.}
	\label{fig1}	
\end{figure}

{\bf Reentrant phase transition regime ($0.3536 < b < 0.5$)} \quad In this parameter range,by solving the Eq.~(\ref{eq:critical value}), the system has two critical points and exhibits reentrant phase transition phenomena. As shown in the left diagram in Fig.~\ref{fig1}, when the pressure is fixed at $P = 0.98P_z$ ($P_z$ corresponds to the pressure at the onset of the zeroth-order phase transition), as the temperature decreases to $T_1$, the system undergoes a first-order phase transition from large black hole to small black hole. The Gibbs free energy of the two phases is continuous, but its first derivative is discontinuous, corresponding to the characteristics of a standard first-order phase transition. This phase transition process is depicted by a orange solid line in the right diagram in Fig.~\ref{fig1}, showing a typical gas-liquid-type coexistence curve and ending at a critical point 1 $(P_{\rm c}^{(1)},T_{\rm c}^{(1)})$, where the heat capacity $C_P$ and isothermal compressibility $\kappa_T$ exhibit divergent behavior. When the temperature is further lowered to $T_0 < T_1$, the system undergoes a transition from small black hole to large black hole, with the Gibbs free energy exhibiting a discontinuous jump $\Delta G \neq 0$ at temperature $T_0$, marking the occurrence of a zeroth-order phase transition. This zeroth-order phase transition starts at the point $(P_z, T_z)$ and ends at the triple point $(P_t, T_t)$. This phase transition region is extremely narrow, and we have determined its corresponding red solid line in the right diagram in Fig.~\ref{fig1}. It is worth noting that this red line is not a traditional phase coexistence curve. Along this line, small black holes have a lower Gibbs free energy and remain in a locally stable state rather than coexisting phases; accordingly, the endpoint of this curve does not satisfy the critical point criterion. In a word, the first-order phase transition at high temperatures and the zeroth-order phase transition at low temperatures together form a complete reentrant phase diagram, profoundly reflecting the nontrivial coupling effects of BI nonlinear electrodynamics with the AdS background \cite{Rasheed:1997ns}.

\begin{figure}[htbp]
	\centering
		\includegraphics[width=70 mm]{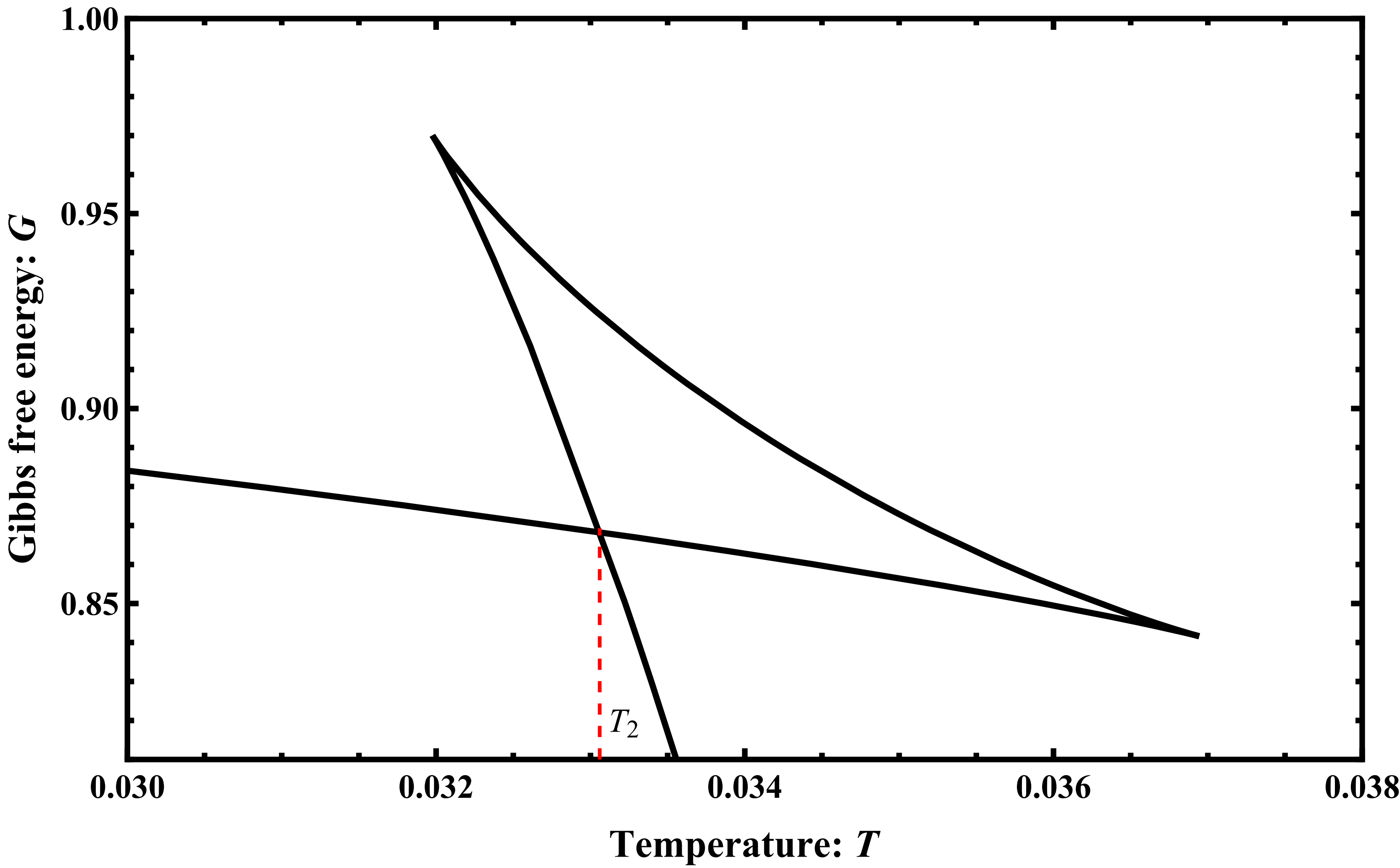}
        \includegraphics[width=70 mm]{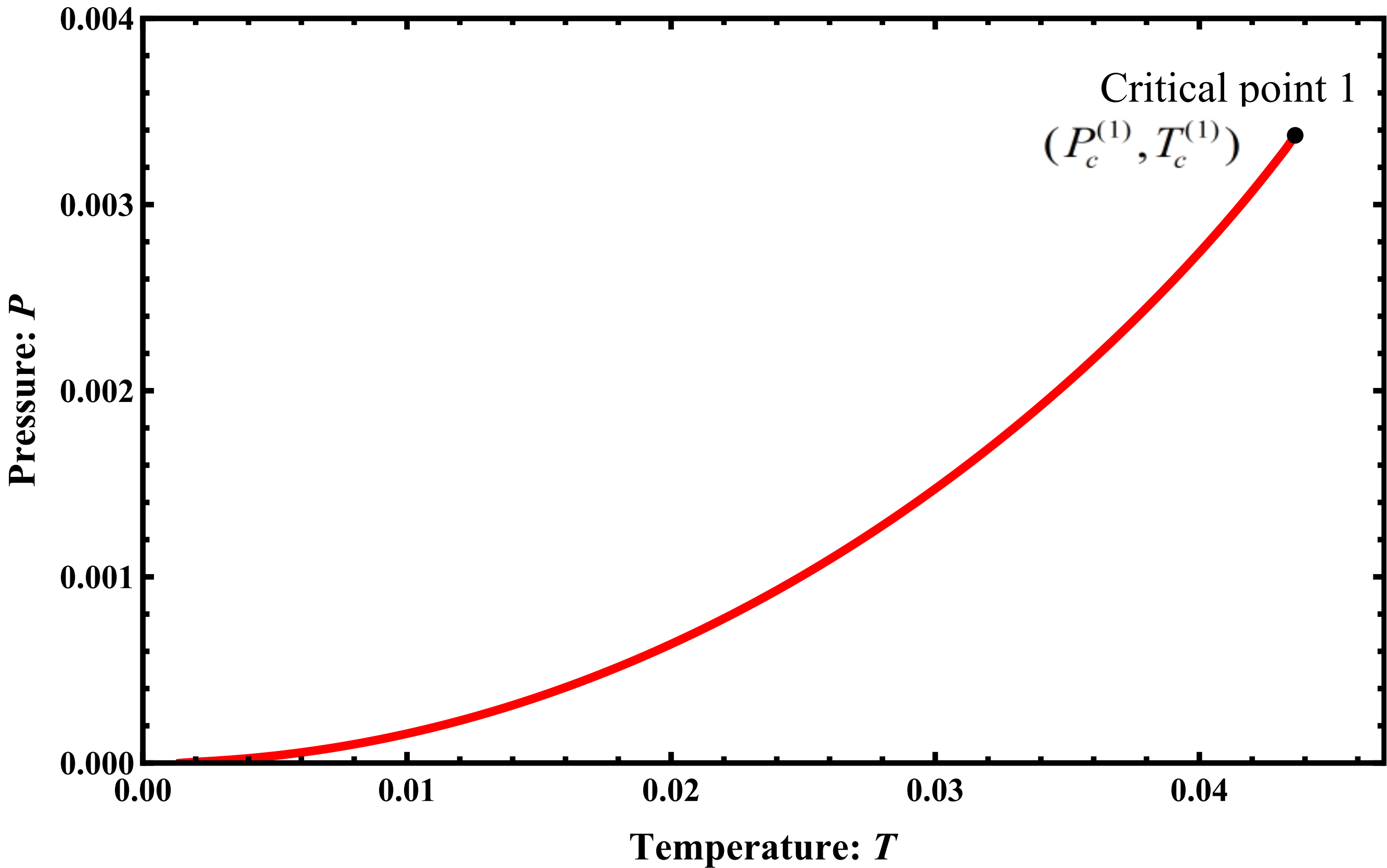}
	\caption{Phase diagrams of BI-AdS black hole with $b=1.0$ and in the left diagram we set  $P=0.5P_c$.}
	\label{fig2}	
\end{figure}

{\bf Standard first-order phase transition regime ($b > 0.5$)} \quad In this parameter range, nonlinear effects become very weak, and the system behavior approaches that of linear Maxwell theory. As shown in Fig.~\ref{fig2},by solving the Eq.~(\ref{eq:critical value}), the system has only one physical critical point. The the left diagram in Fig.~\ref{fig2} exhibits standard swallowtail behavior when the temperature decreases to $T_2$, the black hole undergoes a phase transition from small black hole to large black hole, displaying the typical first-order phase transition behavior analogous to that of four-dimensional RN-AdS black holes \cite{Kubiznak:2012wp}.
	
The above analysis establishes the fundamental thermodynamic framework for four dimensional BI-AdS black holes and classifies the phase transition behavior below the critical point. In the supercritical region of black holes, although conventional phase transitions do not occur, the thermodynamic system still undergoes a continuous transformation from a “liquidlike” phase to a “gaslike” phase, and its boundaries can be precisely determined using Lee-Yang phase transition theory. Next, we will focus on studying its supercritical behavior, and through precise numerical calculations for three typical parameters with $b = 0.4$, $b = 0.45$, and $b = 1$, the geometric structure of the Widom line will be systematically constructed, forming a complete complex phase diagram.
	   	
\section{Complex Phase diagram and Widom line}\label{3}
The core idea of the Lee-Yang phase transition theory is that the essence of a phase transition originates from the distribution of zeros of the partition function (namely the Lee-Yang zeros) in the complex parameter plane. In the thermodynamic limit, when these zeros approach the real axis and form a continuous distribution on it, the derivatives of thermodynamic functions (such as free energy) become non-analytic, and the system undergoes a genuine phase transition. In black hole thermodynamics, the Gibbs free energy $G$ of a black hole can be obtained by calculating the Euclidean action $I_E$, with the relationship between them given by $G=T I_E$. The thermodynamic partition function under the saddle-point approximation can be expressed as $Z = e^{-I_E}$, from which we can derive $G = -T \ln Z$. Hence, it is clear that the zeros of the partition function correspond to the singularities of the Gibbs free energy function (excluding extreme black holes) \cite{Xu:2025jrk}.
	   	
In our analysis, we analytically extend the radius of the event horizon of a black hole to the complex domain $r_h=\text{Re}(r_h)+i~\text{Im}(r_h)$ to obtain the singularities of the Gibbs free energy at arbitrary BI parameters. We find that the second derivative of free energy Eq.~(\ref{eq:fe}) with respect to temperature exhibits a singularity, which coincides with the divergence of the isobaric heat capacity Eq.~(\ref{eq:Cp}), satisfying the following equation, that is, the Lee-Yang zeros distribution equation
\begin{equation}\label{fenbu1}	
2 - 2b^2 r_h^4 - \left(r_h^2 - 2b^2 r_h^4 - 8\pi P r_h^4\right) A = 0.
\end{equation}
Then we will examine the distribution of zeros based on the above equation and derive the complex phase diagram and its supercritical crossover of the four dimensional BI-AdS black hole.

\subsection{Case of $b=0.4$}	
\begin{figure}[htbp]
\centering
\includegraphics[width=120mm]{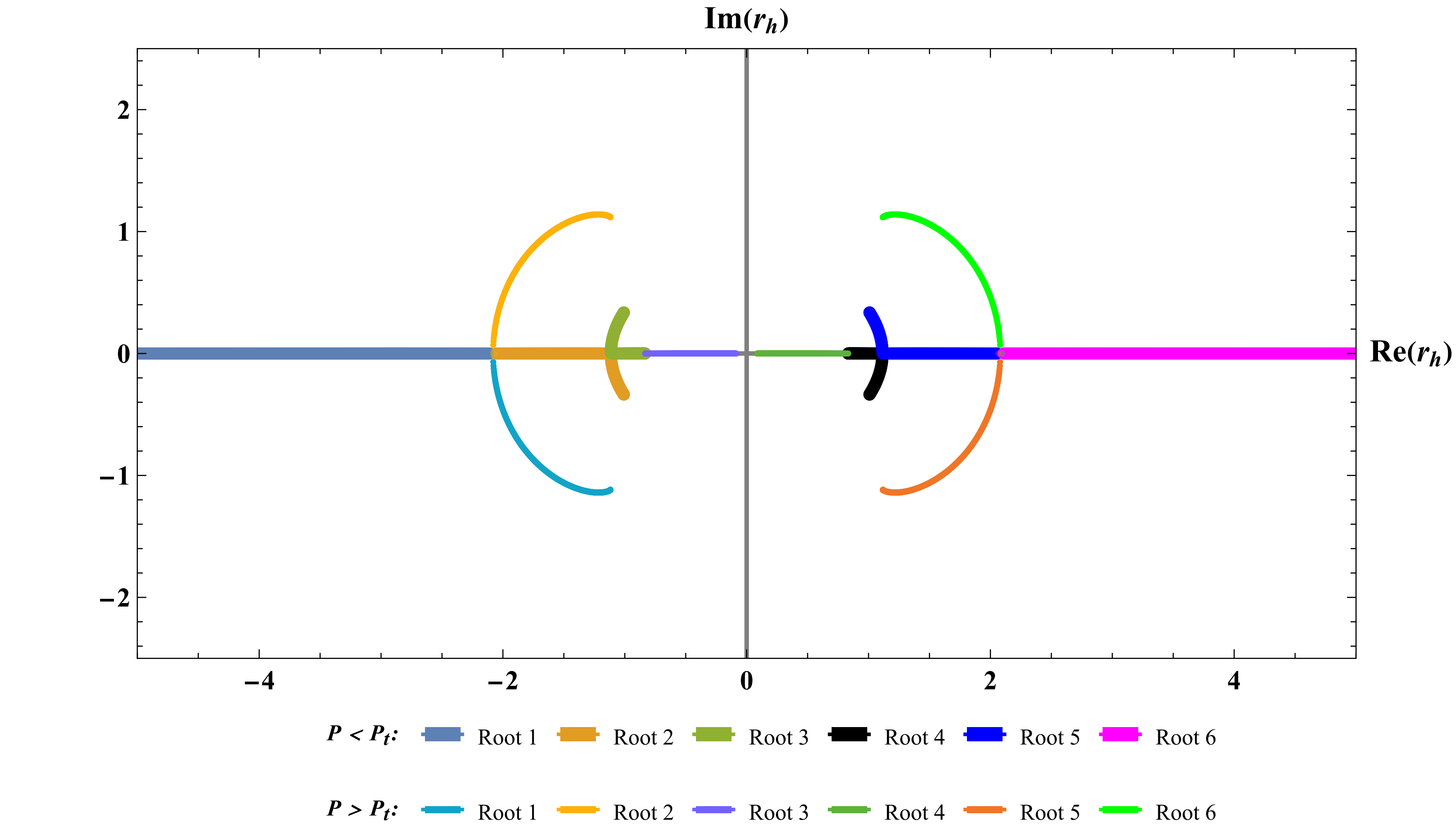}
\caption{Singularities distribution of the Gibbs free energy for the four dimensional BI-AdS black hole at $b=0.4$.}
\label{Fig6}
\end{figure}

When the $\text{BI}$ parameter is set to $b=0.4$, the system corresponds to the reentrant phase transition region in Sec.~\ref{2.2}, as shown in Fig.~\ref{fig1}. In this case, according to Eq.~(\ref{fenbu1}), we can obtain the distribution of Lee-Yang zeros in Fig.~\ref{Fig6}, where we can see that zeros are divided into two categories: (i) $P < P_t$ corresponds to the zeros on the real axis related to phase transitions in the critical region, while (ii) $P > P_t$ corresponds to the zeros distributed in the complex plane related to ``phase transitions'' in the supercritical region. It should be noted that the physically meaningful zeros correspond to those in the first quadrant and on the positive real axis in Fig.~\ref{Fig6}, where both the real and imaginary parts of the event horizon radius are positive.

\begin{figure}[htbp]
	\centering
	\includegraphics[width=170mm]{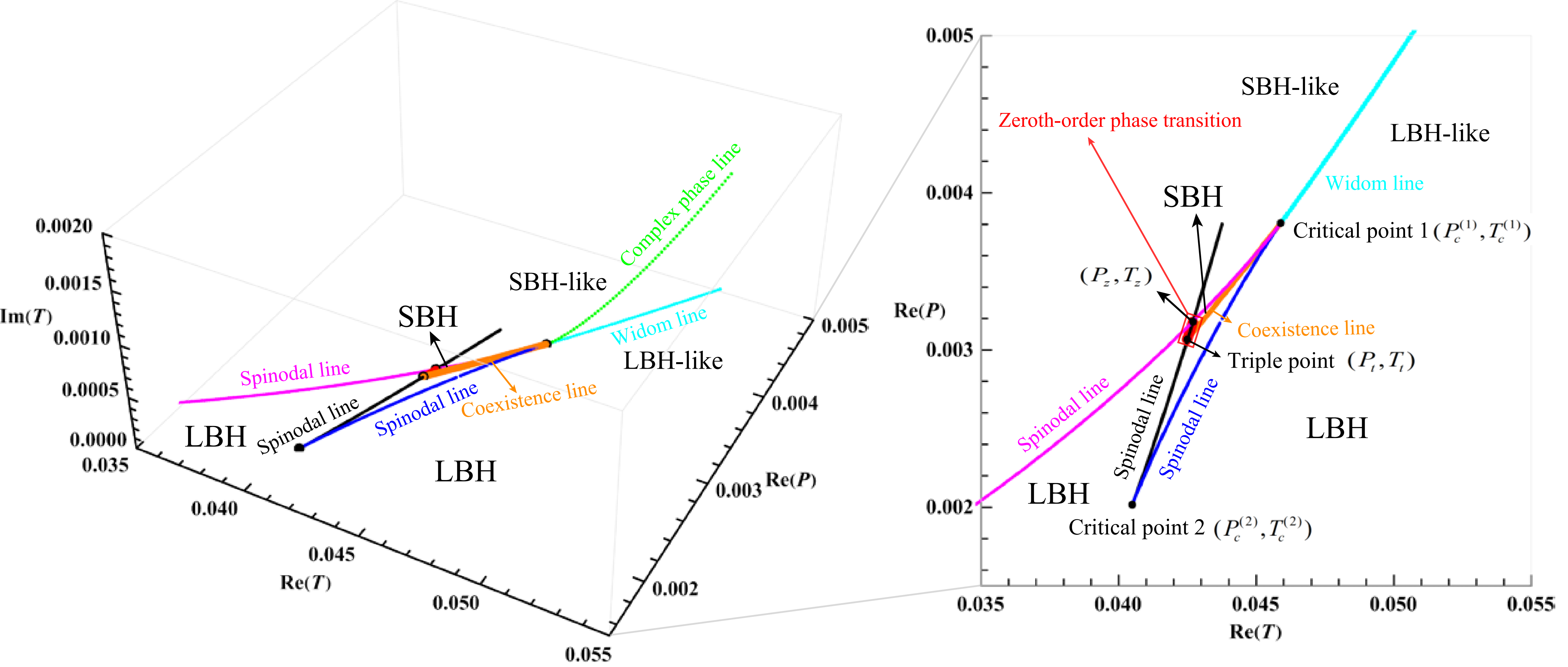}
	\caption{Complex phase diagram of BI-AdS black hole at $b=0.4$.}
	\label{Fig7}
\end{figure}

Correspondingly, the complex phase diagram for the black hole system is depicted in Fig.~\ref{Fig7}. From this, we can obtain the following information.

\begin{itemize}
\item The coexistence line (corresponding to the orange line in the right diagram in Fig.~\ref{fig1}) is obtained through numerical calculations based on Maxwell's equal-area law~\cite{Spallucci2013}. It originates from the critical point 1 $(P_{\rm c}^{(1)}, T_{\rm c}^{(1)})$ and terminates at the triple point ($P_{t}$, $T_{t}$). Traversing this line induces a phase transition from large black hole to small black hole, which is a first-order phase transition. It is worth noting that this phase transition process is accompanied by an additional zero-order phase transition. Similarly, we can also obtain the zeroth-order phase transition ``coexistence line'' (corresponding to the red line in the right diagram in Fig.~\ref{fig1}) through numerical calculations based on Maxwell's equal-area law. It commences at the point ($P_{z}$, $T_{z}$) and also terminates at the triple point ($P_{t}$, $T_{t}$). Traversing this red line triggers a phase transition from small black hole to large black hole. Both types of transitions are characterized by distinct discontinuities in thermodynamic state functions.
    Notably, the ``coexistence curve'' of the zeroth-order phase transition coincides with the black spinodal line. To distinguish, we use a bold red line to represent the coexistence line of the zeroth-order phase transition in the right diagram in Fig.~\ref{Fig7}.
\item The spinodal lines (blue, black and magenta lines) corresponds to the distribution of zeros with the same color in Fig.~\ref{Fig6} for $P < P_t$: the blue and black spinodal lines intersect at critical point 2 $(P_{\rm c}^{(2)}, T_{\rm c}^{(2)})$; the intersection of the magenta and black spinodal lines corresponds to the initiation point of the zeroth-order phase transition ($P_{z}$, $T_{z}$); and the magenta and blue spinodal lines intersect at critical point 1 $(P_{\rm c}^{(1)}, T_{\rm c}^{(1)})$. These spinodal lines are the metastable boundaries.
\item The complex line (green line) corresponding to the distribution of zeros with the green in Fig.~\ref{Fig6}). It extends from the critical point 1 $( P_{\rm c}^{(1)}, T_{\rm c}^{(1)})$ into the supercritical region. This line reflects the phase transition in the supercritical region.
\item The Widom line as the projection of the complex line onto the real plane based on the spirit of~\cite{Xu:2025jrk}, it marks the crossover line of the phase transition between large-like and small-like black holes in the supercritical region.
\end{itemize}

\subsection{Case of $b=0.45$}
\begin{figure}[htbp]
\centering
\includegraphics[width=120mm]{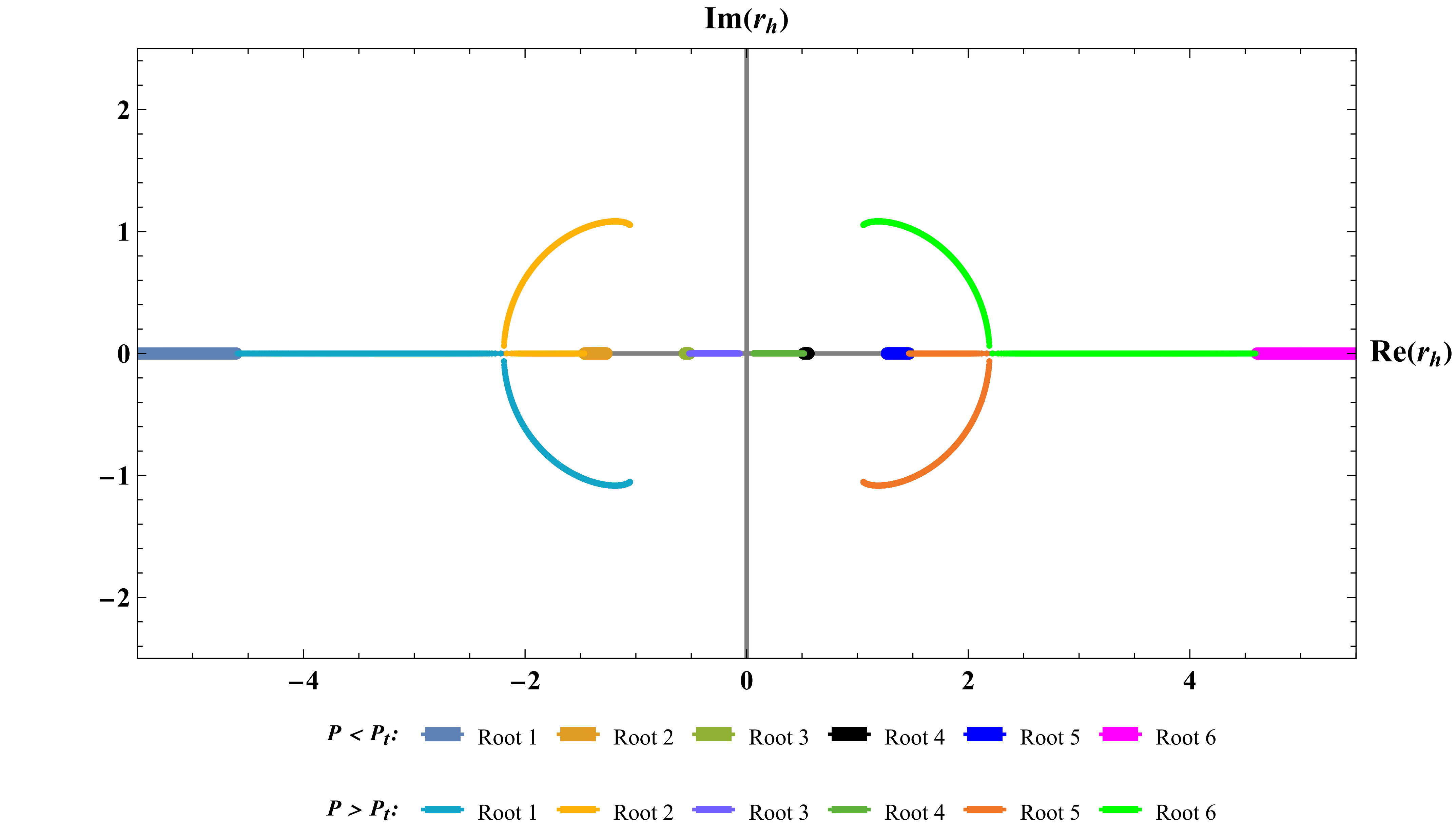}
\caption{Singularities distribution of the Gibbs free energy for the four dimensional BI-AdS black hole at $b=0.45$.}
\label{Fig9}
\end{figure}

\begin{figure}[htbp]
\centering
\includegraphics[width=170mm]{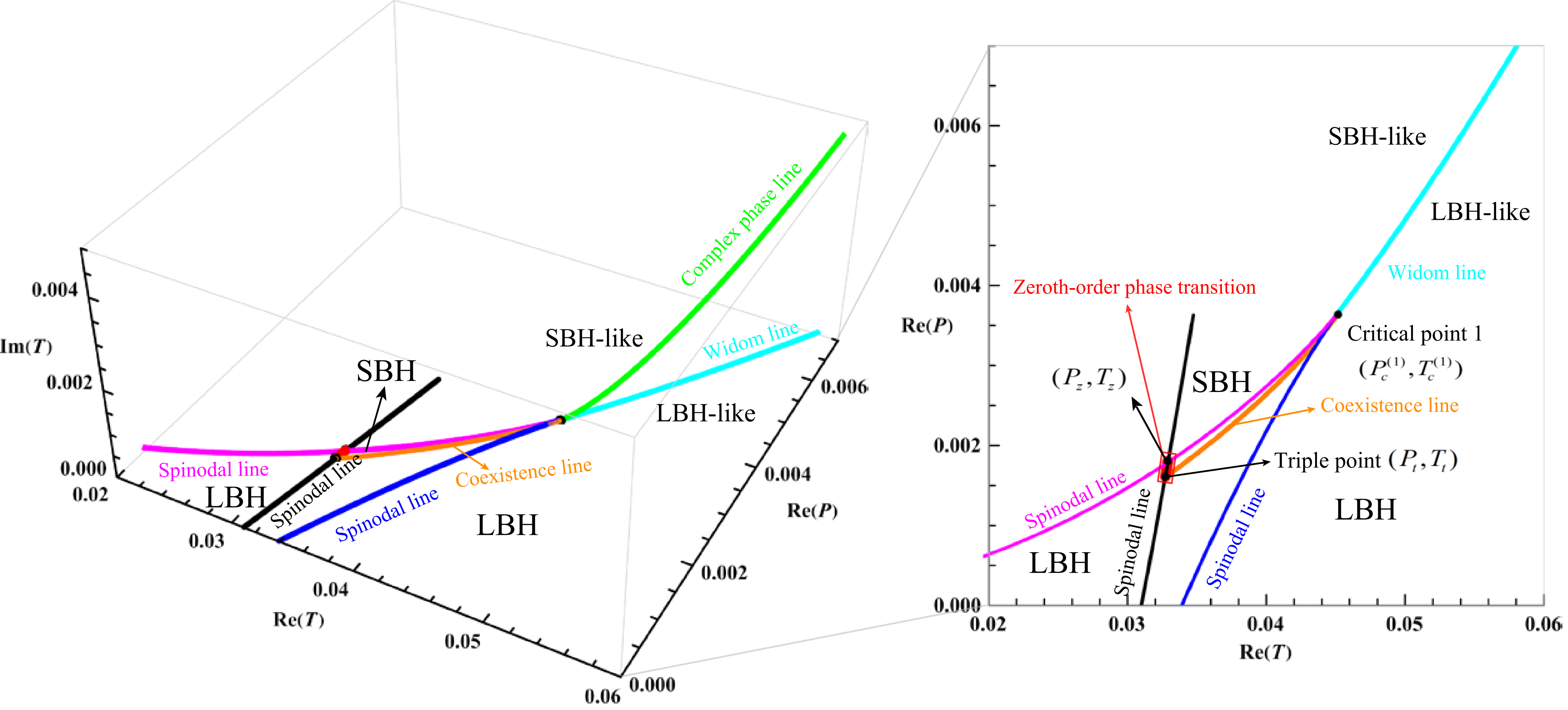}
\caption{Complex phase diagram of the four dimensional BI-AdS black hole at $b=0.45$.}
\label{Fig10}
\end{figure}

To make the results more general, we chose a $\text{BI}$ parameter of $b=0.45$, which still corresponds to the reentrant phase transition region in Sec.~\ref{2.2}. According to Eq.~(\ref{fenbu1}), we can obtain the distribution of Lee-Yang zeros in Fig.~\ref{Fig9}. Correspondingly, the complex phase diagram for the black hole system is depicted in Fig.~\ref{Fig10}.
	   	
By comparing Fig.~\ref{Fig7} and Fig.\ref{Fig10}, we observe that the thermodynamic behaviors for the parameters $b = 0.45$ and $b = 0.4$ are overall identical. The only difference lies in that under this parameter, the intersection point (i.e., critical point 2 $(P_{\rm c}^{(2)}, T_{\rm c}^{(2)})$) of the black spinodal line and the blue spinodal line is non-physical, as the pressure is negative at this point. Under this parameter, a single Widom line still exists only in its supercritical region, extending from critical point 1 $(P_{\rm c}^{(1)}, T_{\rm c}^{(1)})$ into the supercritical region and marking the phase transition in the supercritical region and no Widom line is observed in the region above the ``critical point'' ($P_{z}$, $T_{z}$) of the zeroth-order phase transition.   	
	   	
\subsection{Case of $b=1$}
\begin{figure}[htbp]
\centering
\includegraphics[width=120mm]{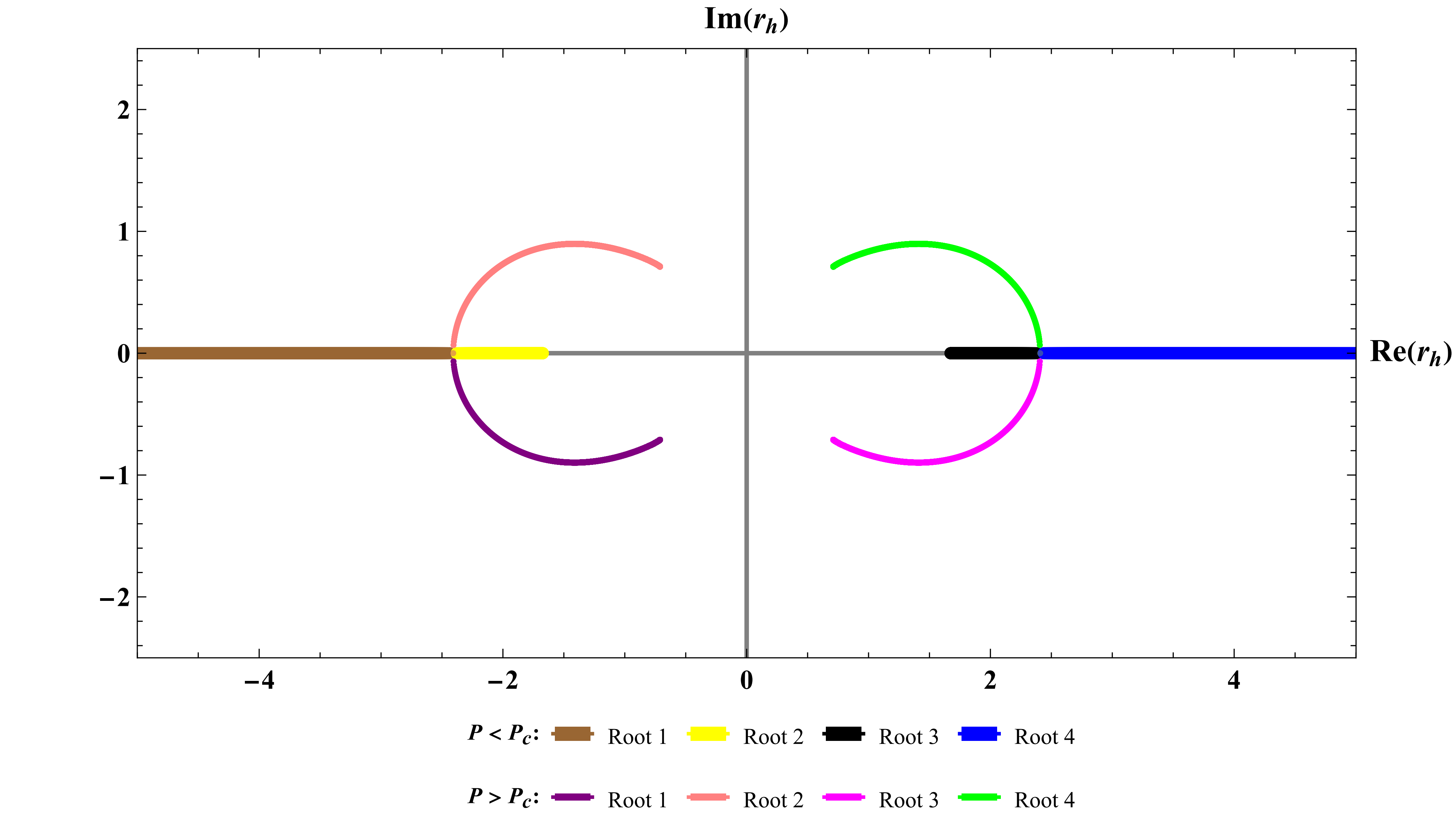}
\caption{Singularities distribution of the Gibbs free energy for the four dimensional BI-AdS black hole at $b=1$.}
\label{Fig12}
\end{figure}

\begin{figure}[htbp]
\centering
\includegraphics[width=170mm]{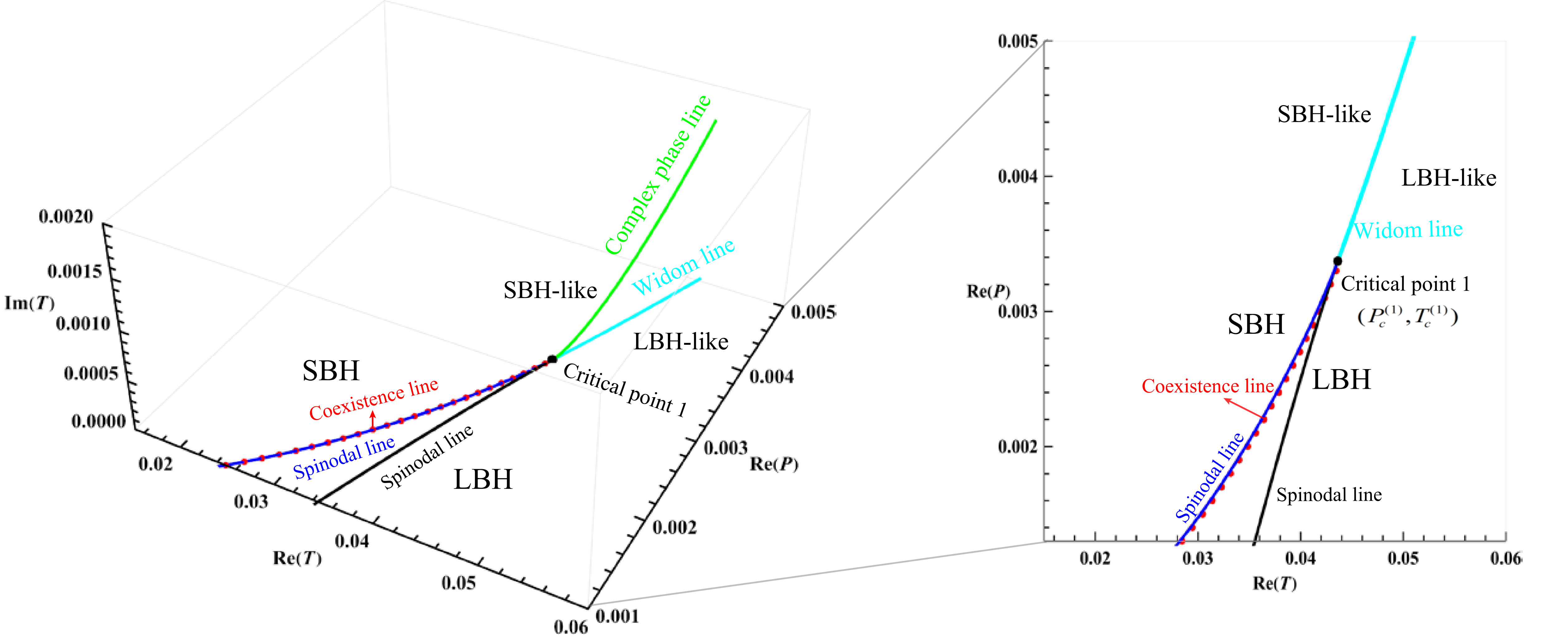}
\caption{Complex phase diagram of BI-AdS black hole at $b=1$.}
\label{Fig13}
\end{figure}

When the $\text{BI}$ parameter is set to $b=1$, the system corresponds to the standard first-order phase transition region in Sec.~\ref{2.2}, as shown in Fig.~\ref{fig2}. In this case, according to Eq.~(\ref{fenbu1}), we can obtain the distribution of Lee-Yang zeros in Fig.~\ref{Fig12}, where we can see that zeros are divided into two categories: (i) $P < P_c$ corresponds to the zeros on the real axis related to phase transitions in the critical region, while (ii) $P > P_c$ corresponds to the zeros distributed in the complex plane related to phase transitions in the supercritical region. Correspondingly, the complex phase diagram for the black hole system is depicted in Fig.~\ref{Fig13}. From the two figures, we can obtain the following information.
	   	
\begin{itemize}
\item The coexistence line (red line) corresponding to the red line in the right diagram in Fig.~\ref{fig2}) is obtained based on Maxwell's equal-area law. It originates from the zero point and terminates at the critical point 1 $( P_{\rm c}^{(1)},T_{\rm c}^{(1)})$. Traversing this line induces a phase transition from large black hole to small black hole, which is identified as a first-order phase transition.
\item The spinodal lines (blue and black lines) corresponds to the distribution of zeros with the same color in Fig.~\ref{Fig12} for $P < P_c$. These spinodal lines are the metastable boundaries.
\item The complex line (green line) corresponding to the distribution of zeros with the green line in Fig.~\ref{Fig12}). It extends from the critical point 1 $(P_{\rm c}^{(1)},T_{\rm c}^{(1)})$ into the supercritical region.
\item The Widom line is the projection of the complex line onto the real plane. It is the boundary between two phases with different physical properties in the supercritical region, according to the idea presented in ~\cite{Xu:2025jrk}.
\end{itemize}
	
In this case, the complex phase diagram of the BI-AdS black hole is similar with that of the four-dimensional charged AdS black hole~\cite{Xu:2025jrk}. It is worth emphasizing that for higher dimensional ($d \geq 5$) BI-AdS black holes, only a standard first-order phase transition is present in the critical region. It can be inferred that their complex phase diagram is also completely similar to that of the four-dimensional case with the parameter $b = 1$.
	
Based on the numerical calculations and analyses presented above, for BI-AdS black holes, we obtain a single Widom line in the supercritical region. This line originates from the stable critical point of the first-order phase transition (critical point 1 $(P_{\rm c}^{(1)},T_{\rm c}^{(1)})$ ) and extends into the supercritical region. While different BI parameter values lead to distinct phase transition behaviors in the critical region, it do not alter the structural characteristics of the supercritical region of the black holes.	
	
\section{Conclusion and Disscusion}\label{V}
Based on the Lee-Yang theory of phase transitions applied in the black hole thermodynamics~\cite{Xu:2025jrk}, we have constructed a complete complex phase diagram for the BI-AdS black holes. This phase diagram integrates the thermodynamic behavior both in the critical and supercritical region, thereby providing a clear and unified representation of the thermodynamic structure of BI-AdS black holes.
	   	
This study concentrates on analyzing the thermodynamic properties associated with the parameters $b=0.4$ and $b=0.45$, both of which lie within the reentrant phase transition regime, as part of an investigation across three representative BI parameter values. The thermodynamic behaviors are similar in the critical region for these two parameters. The sole distinction lies in the pressure at critical point 2 for $b=0.45$, which is unphysical. In addition, the case of $b=1$ is characterized by a standard first-order phase transition. This is established through the construction of its complex phase diagram, which demonstrates that its thermodynamic structure is analogous to that of four-dimensional charged AdS black hole.
	   	
	   	
Notably, although rich phase structures, including reentrant phase transitions, can emerge in the critical region, the supercritical region consistently exhibits only a single Widom line. This line originates exclusively from the stable critical point of the first-order phase transition and extends continuously into the supercritical region. In contrast, no analogous Widom line structure is generated in the case of a zeroth phase transition. When the temperature drops to that of a zeroth-order phase transition, the system experiences a discontinuous jump from the small black hole phase directly to the large black hole phase, without passing through a regime of two-phase coexistence. Given the thermodynamic stability of the small black hole phase, as determined by its lower Gibbs free energy, the associated ``coexistence line'' describes merely a single-phase locus. We refer to the critical point of the zeroth-order phase transition ($P_{z}$ , $T_{z}$) as a pseudo-critical point. It therefore lacks the converging two-phase boundary required for a continuous critical point in the conventional thermodynamic sense.
	   	
Physically, the emergence of the Widom line is fundamentally tied to the long-range fluctuation behavior near a critical point. At the stable critical point of a first-order phase transition, the correlation length diverges and critical fluctuations become pronounced. Hence the Lee-Yang zeros in the complex plane organize into a continuous singular trajectory that emanates from this critical point and extends into the supercritical region. The projection of this trajectory onto the real axis defines the Widom line. In contrast, a zeroth-order phase transition is a discontinuous jump whose transition point exhibits no divergence of the correlation length and therefore inherently lacks the critical fluctuations required to sustain a continuous distribution of singularities. Consequently, such a point cannot give rise to a singular trajectory that extends into the supercritical region and, cannot support a Widom line.

\section*{Acknowledgments}
This research was supported in part by the National Natural Science Foundation of China (Grant No. 12575064, No. 12247103), by the Natural Science Basic Research Plan in Shaanxi Province of China (Grant No. 2025JC-YBQN-029), and by the project of Tang Scholar in Northwest University.

\end{document}